\documentclass[12pt,preprint]{aastex}

\newcommand\gsim{\mathrel{\hbox{\rlap{\hbox{\lower4pt\hbox{$\sim$}}}\hbox{$>$}}}}

\shorttitle{Galaxy Luminosity Profiles}
\shortauthors{Coenda et al.}

\begin{document}

\title{Photometric observations of Southern Abell Cluster Redshifts Survey Clusters: Structure of galaxies in the inner region of clusters of galaxies}

\author{
Valeria Coenda\altaffilmark{1},
Hernan Muriel\altaffilmark{1},
Carlos Jos\'e Donzelli\altaffilmark{1}}
\affil{Grupo de Investigaciones en Astronom\'\i a Te\'orica y Experimental,
IATE, Observatorio Astron\'omico, Universidad Nacional de
C\'ordoba, Laprida 854, X5000BGR, C\'ordoba, Argentina.}
\email{vcoenda@oac.uncor.edu, hernan@oac.uncor.edu, charly@oac.uncor.edu}
\author{Hernan Quintana,
Leopoldo Infante}
\affil{Departamento de Astronom\'\i a y Astrof\'\i sica,
Pontificia Universidad Cat\'olica, Vicu\~na Mackenna 4860, Casilla 306
Santiago 22, Chile.}
\email{hquintana@astro.puc.cl, linfante@astro.puc.cl}
\and
\author{Diego Garc\'{\i}a Lambas\altaffilmark{1}}
\affil{Grupo de Investigaciones en Astronom\'\i a Te\'orica y Experimental,
IATE, Observatorio Astron\'omico, Universidad Nacional de
C\'ordoba, Laprida 854, X5000BGR, C\'ordoba, Argentina.}
\email{dgl@oac.uncor.edu}
\altaffiltext{1}
{Consejo Nacional de Investigaciones Cient\'\i ficas y T\'ecnicas (CONICET),
Avenida Rivadavia 1917, C1033AAJ, Buenos Aires, Argentina.}

\begin{abstract}
We analyze photometric properties of 1384 cluster galaxies
as a function of the normalized distance to cluster center.
These galaxies were selected 
in the central region ($r/r_{200} \leq$ 0.8)  of 14 southern Abell clusters chosen 
from the Southern Abell Cluster Redshifts Survey (SARS).
For 507 of these galaxies we also obtained their luminosity profiles.
We have studied the morphology-clustercentric distance relation
on the basis of the shape parameter $n$ of the S\'ersic's law. 
We also have analyzed the presence of a possible segregation in magnitude for 
both, the galaxy total luminosity and that of their components (i.e. the bulge and 
the disk). 

Results show a marginal ($2\sigma$ level) decrease of the total luminosity as a 
function of normalized radius. However, when bulges are analyzed separately, a 
significant luminosity segregation is found ($3\sigma$ and $2\sigma$ for 
galaxies in projection and member galaxies respectively). The fraction of 
bulges brighter than $M_B \leq -22$ is three times larger in the core of clusters 
than in the outer region.  Our analysis of the disk component suggests that disks are, 
on average, less luminous in the cluster core than at $r/r_{200} \sim 0.8$. 
In addition, we found that the magnitude-size relation as a function of 
$r/r_{200}$ indicates (at $2\sigma$ level) that disks are smaller and centrally 
brighter in the core 
of clusters. However, the Kormendy relation (the bulge magnitude-size relation) 
appears to be independent of environment.
\end{abstract}



\section{Introduction}

It is well known that environment affects galaxy properties as:
morphology, luminosity, color, star formation rate,
gas content and  structure of the subsystems. Different mechanisms have been
proposed to explain how these properties can be affected by the environment.
A large number of galaxies can
be well represented by two major components, the bulge and the disk.
These two components can be affected in many ways when the galaxy is
moving into the environment of a rich cluster. Moore et al.(1998) follow 
the evolution of disks galaxies in a rich cluster (galaxy harassment) 
and find that the result of close encounters is a transformation from 
disks to spheroids. Fujita and Nagashima (1999) suggested that ram pressure 
stripping (Gunn and Gott 1972, Abadi et al. 1999) increases the bulge to 
disk luminosity ratio $B/D$ of normal spiral galaxies due to the 
suppression of the star formation and hence favoring the transformation 
into earlier Hubble types.

Luminosity segregation was detected 
by Rood \& Turnrose (1968), Quintana (1979), Capelato et al. (1980), Yepes et al. (1991) 
and Kashikawa et al. (1998). Moreover, this effect was detected when 
galaxies are considered through their clustercentric distances or their 
velocity dispersions (most luminous galaxies have smaller velocity 
dispersion) (Rood et al. 1972, Biviano et al. 1992). 
However, luminosity segregation have also found opponents like 
Noonan (1961), Bahcall (1973) and Sarazin (1980), who suggested that 
evidences for luminosity segregation are spurious, and mostly due to 
poor background subtraction. However, his optimized fitting procedure was 
applied to the Coma cluster data with little background data.

In this paper we analyze the relations between galaxy structure and photometric 
parameters vs. cluster's environment. Dom\'{\i}nguez et al. (2001)  found 
that parameters defined as a function of the distance to the cluster centre 
are the most appropriated to represent the morphological segregation of 
galaxies in the inner relaxed region of nearby clusters. Since the present 
work concentrates on that region of clusters ($r<r/r_{200}$),  we analyze the 
galaxy properties as a function of the clustercentric distance normalized to 
$r_{200}$. Redshift 
confirmed members and galaxies seen in projection are analyzed separately.
The goal of this paper is to quantify the environment dependence of 
the structural and photometric properties of galaxies in clusters.
Our sample consists 
of 507 galaxies in 14 southern Abell clusters of the SARS sample 
(Way et al. 2005, Hearafter Paper I). The structural and photometric parameters were 
obtained in Coenda et al. (2005, Hereafter Paper III). The paper is structured as 
follows: observations and photometric analysis are described in section 
\ref{photo}. In section \ref{res} we derive and analyze our results, 
and the conclusions are given in section \ref{concl}.


\section{Sample, observations and luminosity profiles}\label{photo}

Our sample consists of Cousins \textit{R} CCD images of 14 Abell clusters with 
$cz<40000kms^{-1}$ corresponding to the Southern Abell Clusters Redshifts 
Survey (SARS, Paper I). The images were taken with the Swope 1.0 m 
telescope at Las Campanas Observatory, Chile. The pixel scale was 0.61$\arcsec$
and the field covers a 20.8$\arcmin$ square area. The seeing conditions were very similar for the whole sample of clusters. The mean seeing was 2.00 $\arcsec$ and the dispersion 0.18. Therefore, we believe that  our sample is free of any effect due to variations in the seeing conditions. Additional details on the observations and data reduction are given in Paper III. Table \ref{table1} lists 
the cluster sample, their coordinates, velocity dispersions and radial 
velocities taken from Muriel et al. (2002, Hereafter Paper II). From the listed clusters we have finally analyzed a total of 1384 galaxies (CS1 sample) of which 345 have known redshifts (CS2 sample) (Paper I). Of the 345 with measured redshifts, 313 are cluster members (CS3 sample). Of the 1384 galaxies, 507 have luminosity profiles determined (CS4 sample) and 232 of these galaxies have measured redshifts (CS5 sample), and of these 232, 207 are cluster members (CS6 sample).  Luminosity profiles were obtained using the $ellipse$ 
routine within STSDAS (Jedrzejewski, 1987) and were fitted 
using the standard B + D law:
\begin{equation} \label{dV+exp}
I(r)=I_{e}exp\Big[-7.688\Big[\Big(\frac{r}r_{e}\Big)^{1/4}-1\Big]\Big]+I_{0}exp\Big(-\frac{r}r_{0}\Big)
\end{equation}
The first term corresponds to the bulge component, being $I_{e}$ the effective intensity and $r_e$ the effective radius defined as the radius that encloses half of the total luminosity of the bulge. The second term corresponds to the
disk component, being $I_{0}$ the central intensity and $r_{0}$ the length scale.
 In addition, we have also used the S\'ersic law (S\'ersic 1968) 
to fit the galaxy luminosity profiles:
\begin{equation} \label{sersic}
I(r)=I_{s}exp\Big(-\Big(\frac{r}r_{s}\Big)^n\Big)
\end{equation}
In this equation $I_{s}$ is the central intensity and $r_{s}$ the length scale. The exponent $n$ is a shape parameter, where $n=0.25$ correspond to the de Vaucoulueurs law (de Vaucouleurs 1948) and $n=1$ correspond to the exponential law (Freeman 1970). Further details on the fitting procedure and error 
sources can also be seen on Paper III.

\section{Results and Discussion}\label{res}

\subsection{Selection effects}\label{effects}

\begin{figure}
\begin{center}
\includegraphics[scale=0.6]{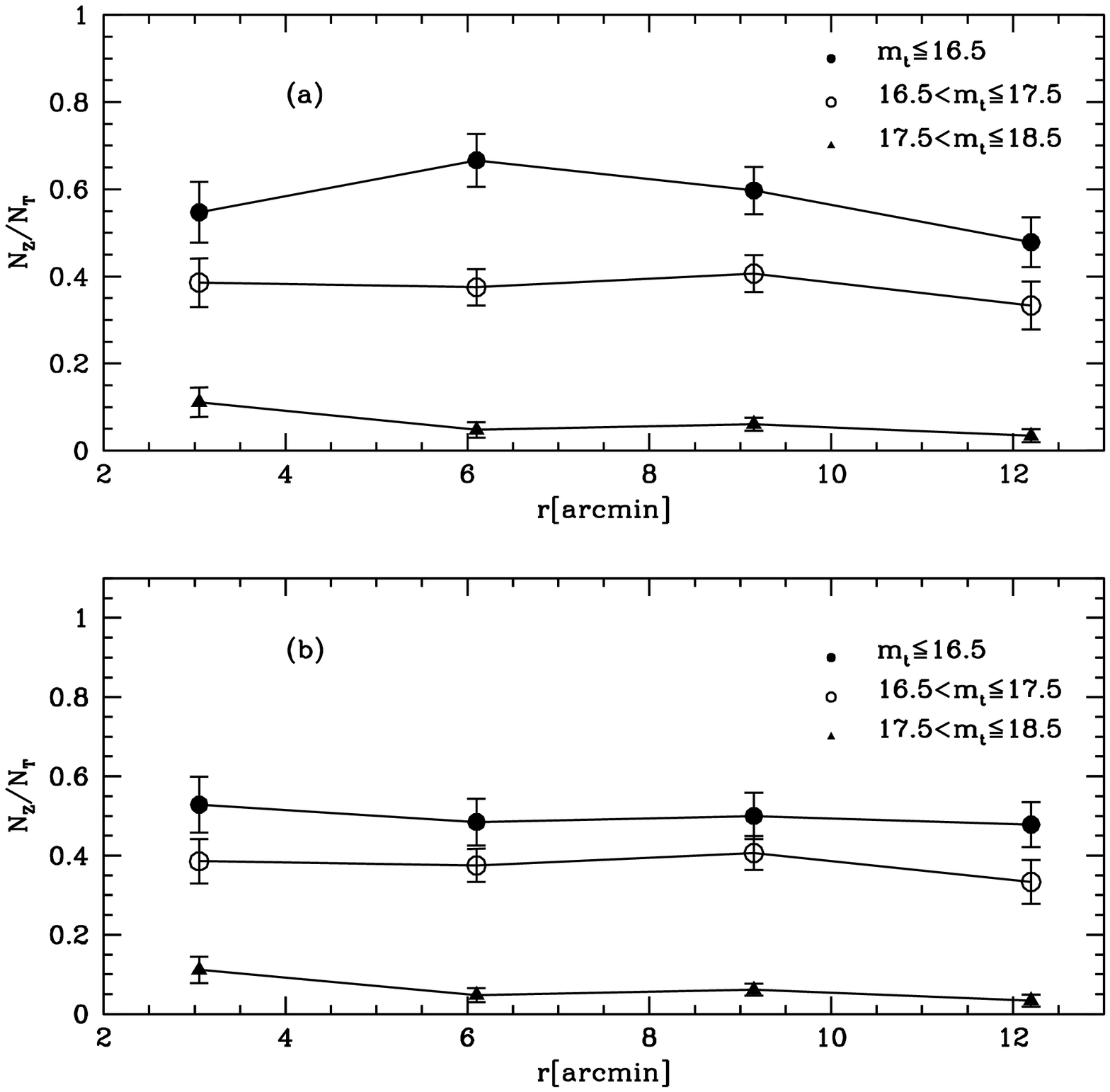}
\caption{(a) Radial distribution of the ratio between the number 
of galaxies with known redshifts and the total number of galaxies with a limit magnitude at $m_t=18.5$, on each of the selected magnitude intervals. (b) Same as (a) but bias corrected.}
\label{fig1}
\end{center}
\end{figure}

In this work we aimed to determine the presence of possible correlations 
between galaxy photometric parameters and cluster global properties. 
Particularly, we focused our study on the morphological and magnitude 
segregation through radial correlations. This is justified from the expetations that different
physical processes are likely to operate at different radii, thus looking for
trends with radius are clearly appropriate if one is trying to understand why we see
luminosity and morphological segregations in clusters. In order to avoid systematic 
effects we have analyzed the 
sample completeness as well as projection and selection effects. To study the
magnitude segregation we must be sure that our sample is free of any radial bias
in the galaxy selection. Since SARS does not represent a magnitude complete 
sample (see Paper I for a more detailed discussion) we have investigated
the possible presence of a radial bias between the SARS target selection and 
apparent magnitude of the galaxies. We  analyzed the radial distribution of 
the quotient between the number of galaxies with known redshifts and the 
total number of galaxies in  a total-magnitude-complete sample (CS1 sample)
(limited at $m_t=18.5$). The analysis was done for three different intervals 
of the total apparent magnitude $m_{t}$ (bulge + disk). As can be appreciated 
in Figure \ref{fig1} (a) the fraction of bright galaxies with measured redshift
present a slight increment towards the central inner region of clusters. 
In order to avoid any possible bias in our analysis, we have randomly selected a new
sample that is free of this bias. We proceeded as follow: randomly selected galaxies with known redshifts were discarded until the fraction of galaxies with redshift as a function of $r$ is nearly constant. This procedure was applied  on each of the selected magnitude intervals. Figure \ref{fig1} (b) shows the resulting galaxy distributions. As a consequence of this procedure we have a new sample with 345 galaxies with measured redshifts (CSC1 sample). 

\begin{figure}
\begin{center}
\includegraphics[scale=0.6]{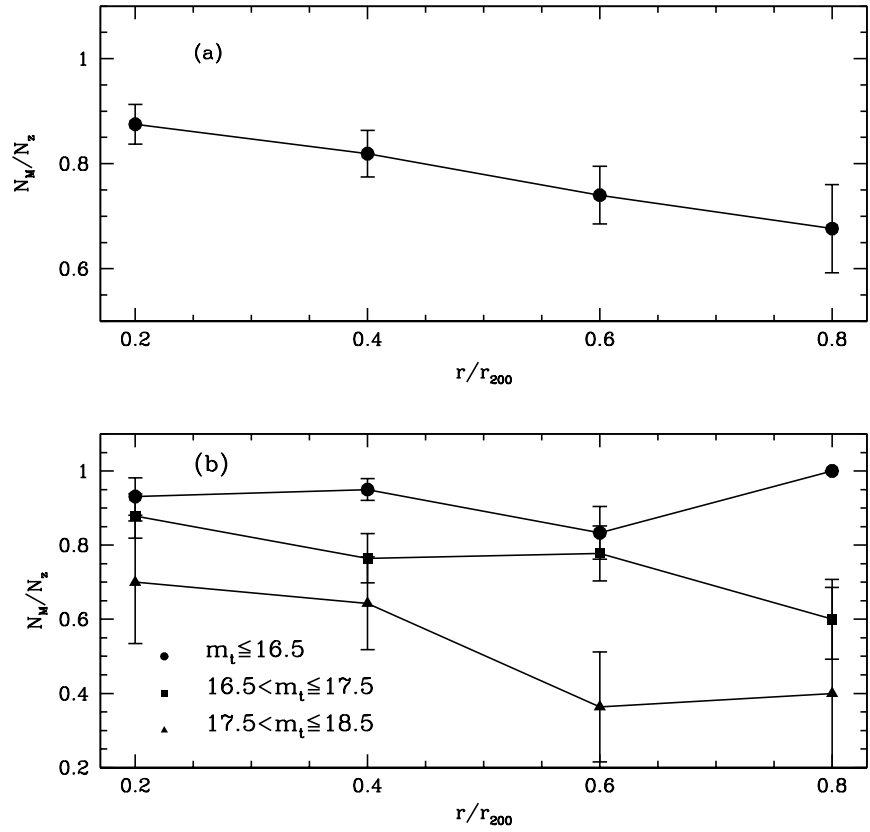}
\caption{Ratio between the number of redshift-confirmed members and the total number of galaxies with known redshifts as a function of $r/r_{200}$ (a) for the total sample and (b) for different magnitude interval.}
\label{fig2}
\end{center}
\end{figure}

Once the the sample is free of any bias in the redshift selection function,
the redshift information can be used to quantify the fraction of galaxies 
that are cluster members as a function of the normalized radius and apparent magnitude.  This information will be used to correct for projection 
effect the sample of galaxies without redshit estimates.  In order to quantify 
this projection effect, and using the sample of galaxies free of the redshift 
selection bias (CSC1 sample), we computed the ratio between
the number of redshift-confirmed members and the total number of galaxies with
known redshift as a function of a normalized radius. This ratio was computed for both, the total sample and for different 
intervals of total apparent magnitude (Figures \ref{fig2} (a) and \ref{fig2} 
(b) respectively). Each galaxy without redshift is weighted depending on  the clustercentric distances and the total apparent magnitude.

The previous analysis was carried out to study the total luminosity segregation.
On the other hand, to study the bulge-disk luminosity segregation 
we performed a similar analysis considering a subsample of galaxies with 
luminosity profiles in which profile decomposition was possible. 

In order to correct for the magnitude limit, we first determined the 
luminosity function of the nearest cluster for both all galaxies 
in projection and for those redshifts-confirmed member galaxies. The obtained
luminosity functions were then fitted with the Schechter's function. 
These fittings were used to correct the observed galaxy 
counts according to Whitmore et al. 1993. The same procedure was applied to 
the analysis of the bulge and the disk systems.

Since we were also interested in the study of the morphological segregation, 
we have analyzed both completeness and projection effects for those galaxies 
where we could trace out the luminosity profile. 
This analysis is similar to the others mentioned above.
However, in this case we have considered different intervals of the S\'ersic profile parameter $n$.  

\subsection{Morphological Segregation}

We have adopted for our analysis the cluster characteristic 
radius, $r_{200}$. This radius is
defined as the radius where the mean inner density is $200\overline{\rho}(z)$. 
Carlberg et al. (1997) derive a correlation 
between $r_{200}$ and the cluster mean velocity dispersion ($\sigma$):
\begin{equation}
r_{200}=\frac{\sqrt{3}\sigma}{10H(z)}
\end{equation}
The values of $\sigma$ used in this work are those quoted in Paper II and the corresponding values of $r_{200}$ can be found in its Table 1.

\begin{figure}
\begin{center}
\includegraphics[scale=0.6]{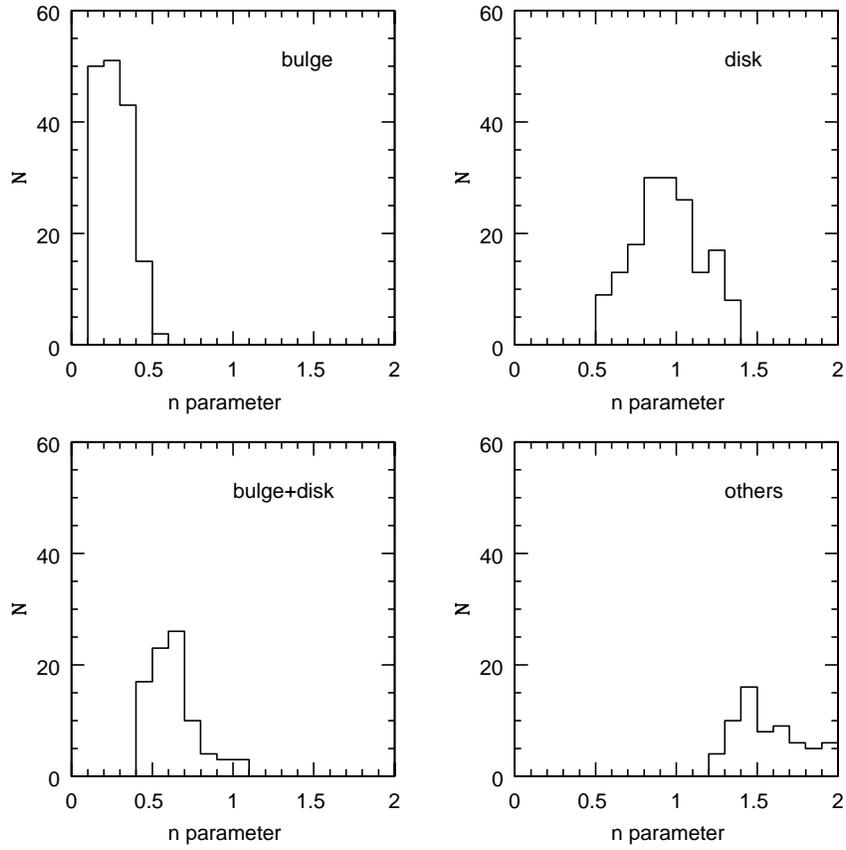}
\caption{Distribution of the $n$ parameter for galaxies having pure de Vaucouleurs profiles, B + D profiles, pure exponential profiles, and galaxies for which we could not fit any of the previous functions.}
\label{fig3}
\end{center}
\end{figure}

As it was pointed out in the introduction, the morphology-environment relation
of galaxies in clusters has been extensively studied (Dressler 1980, Witmore 
et al. 1993, Dominguez et al. 2001). However, in this work we wanted to analyze
the mentioned relation in an alternative way using the $n$ parameter which has 
two major advantages, it is a continuous index and it can be easily reproduced. 
In order to study the possibility of $n$ being a rough morphology indicator we have 
explored the $n$ values fitted to those galaxies with different types of 
luminosity profiles. In Fig. \ref{fig3} we have plotted the $n$ distribution 
for those galaxies having pure de Vaucouleur profiles, B + D profiles and pure 
exponential profiles. The fourth group includes those galaxies for which we 
could not fit any of the previous functions. It is clearly observed that those 
galaxies with r$^{1/4}$ luminosity profiles have values of $n$ $<$ 0.4. On the 
other hand, those galaxies with typical B + D profiles show 0.4 $<$ $n$ $<$ 
0.7, while disk galaxies show 0.7 $<$ $n$ $<$ 1.2. Finally, we have observed 
that those galaxies with 1.2 $<$ $n$ are mostly dwarf ellipticals. 

\begin{figure}
\begin{center}
\includegraphics[scale=0.6]{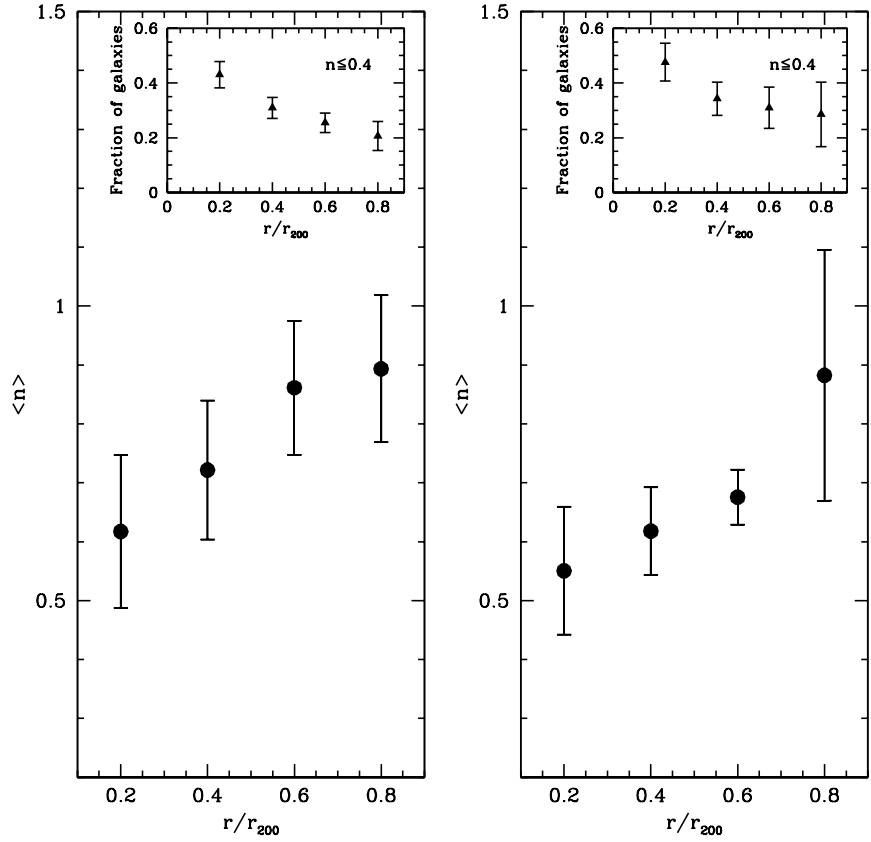}
\caption{Correlation between the average $n$ parameter and normalized clustercentric distance for the sample galaxies, projetion corrected (left panel) and confirmed cluster members (right panel). The small box in the upper corner displays the fraction of galaxies with $n\le0.4$ as function of $r/r_{200}$.}
\label{fig4}
\end{center}
\end{figure}

In figure \ref{fig4} we can observe the correlation between $n$ and $r/r_{200}$. The left panel corresponds to all galaxies in the sample corrected for projection effects whereas the results shown in the right panel are computed  for redshift-confirmed members and corrected for completeness as it was explained in section \ref{effects}. The comparison between the results shown in both panels gives information about the bias introduced by projection effects. As expected, we found that galaxies with low values of $n$, which roughly correspond to early type galaxies, dominate the central cluster region. We also found that the fraction of early type galaxies increases when the clustercentric distance decreases.  Although the behavior in the mentioned panels are similar, the signal is stronger when redshift-confirmed members are considered. Error bars in this and following figures were estimated using the boot-strap re-sampling technique.


\subsection{Segregation in Magnitude}
We are interested in studying a possible magnitude segregation in clusters.
Solanes et al. (1989) used the mean magnitude to test possible dependencies of 
magnitudes as a function of the projected local density. Probably, one of the best 
options is the computation of the luminosity function at different clustercentric 
radii or densities. Nevertheless, this option requires a huge amount of galaxies. 
An alternative option consists in the computation of the fraction of galaxies 
brighter than a certain value as a function of the clustercentric radius. 

\begin{figure}
\begin{center}
\includegraphics[scale=0.6]{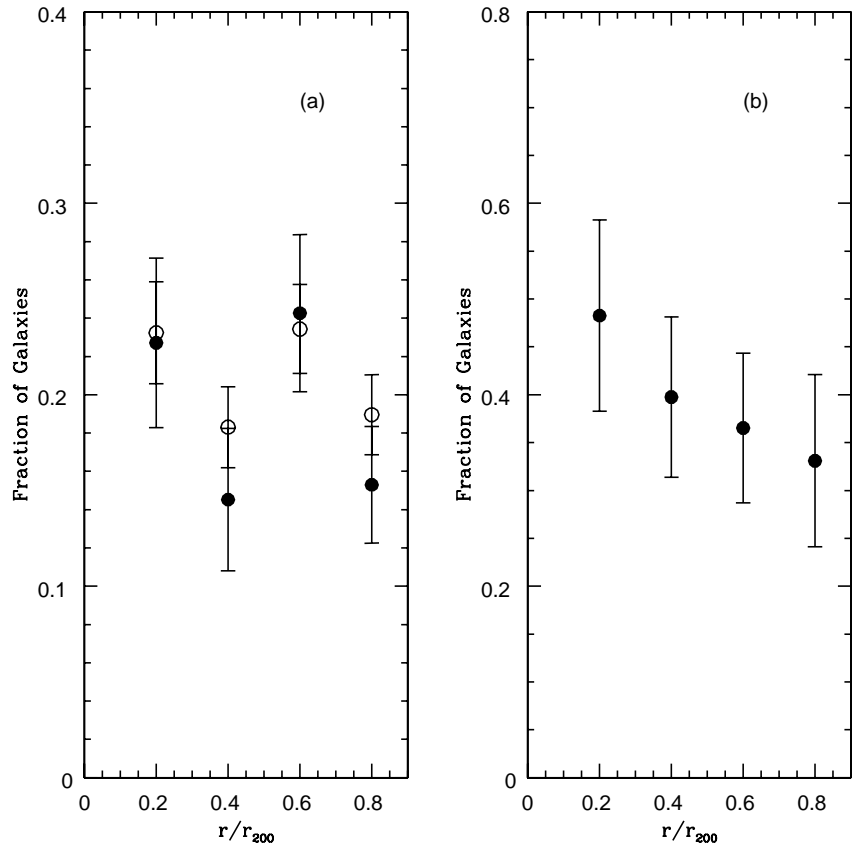}
\caption{Fraction of galaxies with $M_t\le-21$ as function of $r/r_{200}$. (a) for all galaxies of the seven nearest clusters (solid circles) and total sample galaxies (open circles) both corrected for projection effects. (b) for redshift-confirmed members of the seven nearest clusters.}
\label{fig5}
\end{center}
\end{figure}

In order to evaluate a possible segregation in magnitude, we have considered
three possible parameters: the total absolute magnitude, the bulge magnitude
and the disk magnitude. Figure \ref{fig5} plots the fraction of galaxies with
$M_{t}\le-21$ as a function of $r/r_{200}$, using the CS1 sample and CS3 sample.
 The threshold magnitude was selected in order to have two sub-samples of similar size. 
It can be observed that galaxy
total luminosity marginally (two sigma level)decreases as a function of $r/r_{200}$ for
redshift-confirmed members of the nearest seven clusters (panel (b)).
On the other hand, panel (a) shows all galaxies of the previous seven clusters
(solid circles)
and all cluster galaxies (open circles) respectively. In this case, both
samples were
corrected for projection effects. However, we can not see a clear correlation
between
$M_{t}$ and $r/r_{200}$.

\begin{figure}
\begin{center}
\includegraphics[scale=0.6]{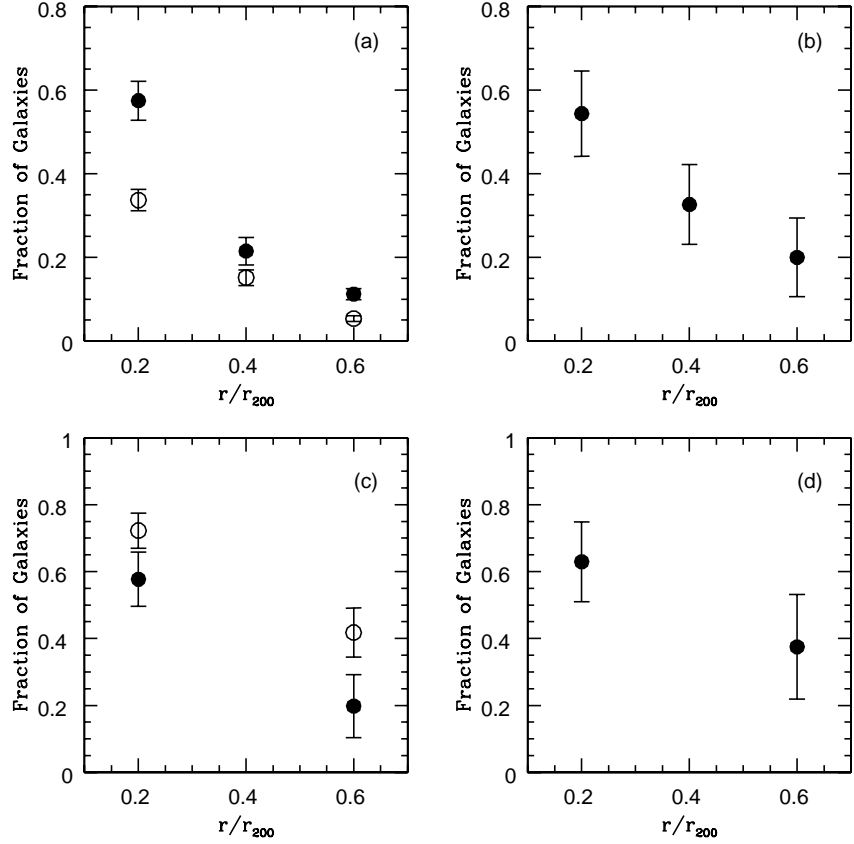}
\caption{Panels (a) and (b) show the fraction of galaxies with $M_B\le-22$ as function of $r/r_{200}$ for the same sample galaxies of Fig. \ref{fig5}. Panels (c) and (d) show the fraction of galaxies with $M_B\le-22$ as function of $r/r_{200}$ but in this case limited to a bulge-magnitude-complete subsamples.}
\label{fig6}
\end{center}
\end{figure}

The study of the bulge sub-system is particularly important since it is a
fundamental component of a high fraction of galaxies and seems to have some
properties that are independent of the morphological type of the host galaxy.
Figure \ref{fig6} shows the fraction of galaxies with $M_{B}\le-22$ versus the
clustercentric distance, panels (a) and (b). We can see that bulge luminosity
decreases as of $r/r_{200}$ increases in all cases.
The sample (CS4 sample and CS6 sample) used in this analysis is complete up to $m_t=18.5$ and therefore
it is not necesary complete in bulge magnitude. For this reason, we repeated
the
previus analysis selecting a bulge-magnitude-complete subsample (CS7 sample). 
Assuming an euclidian geometry and an
uniform space distribution, the function $N_{dot}*10^{0.6m_b}$ provides a good estimate 
of the magnitude completeness. We found that our bulge sample is approximately complete up
to $m_b=16.5$.  The results can be seeing in panels (c) and (d) of Fig.
\ref{fig6}. Analogously, we also
found a clear correlation between bulge luminosity and $r/r_{200}$.
In these last two cases, we have considered only two bins for statistical
reasons.

\begin{figure}
\begin{center}
\includegraphics[scale=0.6]{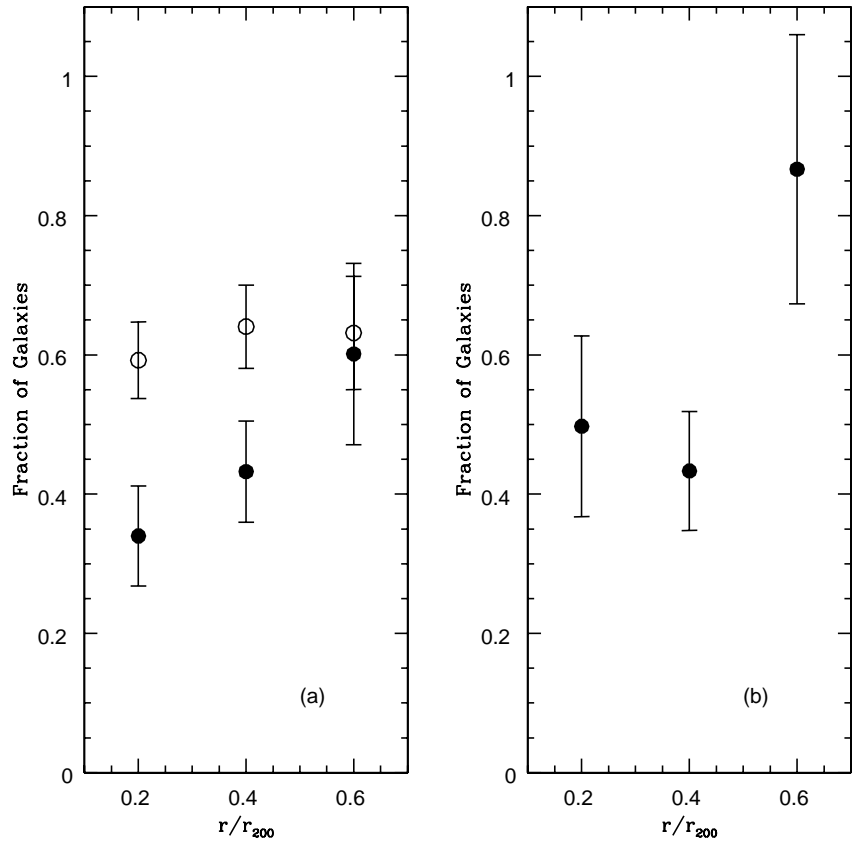}
\caption{Fraction of galaxies with $M_D\le-20.5$ as function of $r/r_{200}$. Panels (a) and (b) correspond to the same cases of Fig. \ref{fig5}.}
\label{fig7}
\end{center}
\end{figure}

As it was discussed in the introduction, disks can be seriously affected by
several processes when they move close to the center of massive clusters of 
galaxies. Figure \ref{fig7} shows, for the same sample of Fig.\ref{fig6},
the fraction of galaxies with $M_{D}\le-20.5$ as a function of $r/r_{200}$. It can be observed 
a clear dependence of $M_{D}$ with the clustercentric distance  (disks become 
brighter as the clustercentric distance increases). It should be noted that 
the effect can not be clearly appreciated when all clusters in   
projection are considered (open circles in panel b).

As it was detailed in Paper III our sample galaxies were selected under a strong restriction: only those galaxies with apparent radius greater than 3-4 times the FWHM were chosen. This selection avoid any bias
in the accuracy of measuring bulge and/or disk luminosities. Several tests described in Paper III show that
the main error source for the photometric parameters is background noise and an eventually dependence on the bulge and/or disk luminosity should not be noticeable in terms of the calculated errors.

\subsection{Scaling Relations}
\begin{figure}
\begin{center}
\includegraphics[scale=0.6]{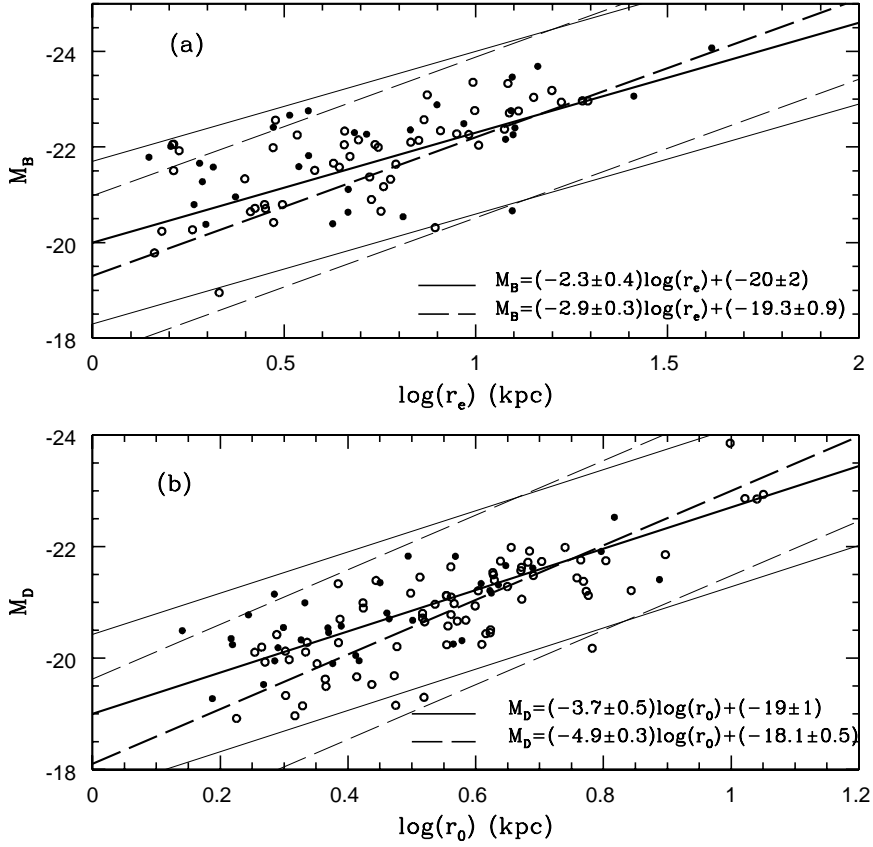}
\caption{$M_B-log(r_e)$ (a) and $M_D-log(r_0)$ (b) relation. In both cases, galaxies with $r/r_{200}<0.3$ are represented with filled circles and the bisector fit is shown with a solid line. Galaxies with $r/r_{200}\geq0.3$ are plotted with open circles and the bisector fit is shown with a dashed line.}
\label{fig8}
\end{center}
\end{figure}

In paper III, we have studied several scaling relations between photometric and
structural parameters. In order to test whether these relations depend on the
cluster environment, we have correlated their behavior as a function of the  
over-density radius. We have studied the  magnitude-size relation, which is 
closely related to the Kormendy relation (Kormendy 1977) for bulge systems. 
Figure \ref{fig8} shows the $M_B-log(r_e)$ (panel a) relation together with 
the 
$M_D-log(r_0)$ relation (panel b) for member galaxies. We have separately 
analyzed 
this scaling relations for those galaxies with $r/r_{200}<0.3$ (filled circles)
and for those with $r/r_{200}\geq0.3$ (open circles). The best fitting 
parameters are shown in the corresponding panels, the slopes and zero points 
correspond to the bisector fit as described in Paper III.  No statistically significant difference between bulges in the inner and in the outskirts part of the clusters was found.
 
On the other hand, our results indicate a marginal dependency (two $\sigma$ level) of the  $M_D-log(r_0)$ relation on environment. We can observe from panel (b) of Fig. 8 that for low values of $log(r_0)$ disks located in the inner cluster regions are brighter than those at intermediate distances from the cluster center. This result implies that the central intensity of disks $I_0$ would be greater for 
galaxies located in the inner cluster region. Nevertheless, more data are required in order to confirm this result.

\section{Conclusions}\label{concl}

We have analyzed the correlation between galaxy photometric parameters
and the normalized clustercentric radius $r/r_{200}$ for 507 galaxies 
in the central region of 14 Abell cluster. All the analysis performed 
in this work were applied to two different samples: i) all galaxies in 
projection, for which we have taken into account the standard corrections, and ii) redshift 
confirmed members corrected by completeness.

Based on the S\'ersic index $n$ we analyzed the morphological-$r/r_{200}$
relation. We found that the $n$ parameter is a good alternative to
measure the morphological segregation. The use of $n$ has the advantage
that it is a continuous parameter that can be estimated in a more reproducible
procedure.

In order to test for a possible luminosity segregation, we analyzed the 
correlation between the fraction of galaxies with $M_{t}\le-21$ and 
$r/r_{200}$. Our results show a marginal (two sigma level) decrease of the 
total luminosity as the normalized radius increases. It should be noted that
this effect is only present when redshift-confirmed members are considered.
The same analysis was repeated for bulge and disk sub-systems.
Our results indicate a segregation in bulge-magnitude when both
confirmed members galaxies ($2\sigma$ level) and projected galaxies 
($3\sigma$ level) are used. 
We found that the fraction of bulges brighter than  $M_{t}\le-22$ is  
approximately three times larger in the inner cluster region than in the 
outer cluster region. This analysis was performed for a total-magnitude-complete 
sample and for a bulge-magnitude-complete sub-sample. In both cases, our results are 
consistent with a segregation in the bulge luminosity. 

On the other hand, the absolute magnitude of disks presents a dependence on 
$r/r_{200}$ in 
the sense that disks tend to have on average lower luminosities as they are 
closer to the core of the parent cluster of galaxies. It should be noted that
this effect is only present when redshift-confirmed members are considered. 

If disk galaxies are selected in projection, the effect is only statistically 
significant for the nearest clusters. 
Disk luminosity segregation support the idea that disks are 
tidally affected by the cluster potential or by high speed encounters with 
other cluster member galaxies. However, there are other phenomena that could affect 
star formation in galaxy disks such as ram pressure, gas evaporation, or lack of gas 
accretion from the intra-cluster medium. All these mechanisms can also be responsible 
for the observed disk luminosity segregation.

We have analyzed the scaling relations for bulges and disks as a function of 
$r/r_{200}$. We did not find any statistically significant dependence of 
the $M_B-log(r_e)$ relation with the clustercentric distance. Our results 
suggest that the physical conditions responsible for the Kormendy relation 
are sufficiently robust to support the extreme conditions found in the core of the 
clusters of galaxies. On the other hand, the $M_D-log(r_0)$ relation is consistent with a marginal 
dependency (two sigma level) of this relation as a function of the 
normalized radius. The  correlation between these two parameters appears 
weaker in the cluster core when compared with 
larger clustercentric distances. This result indicates that disks of galaxies 
in the central region of clusters are more compact and have a brighter
central luminosity. Nevertheless, it should be remembered that, on average, disks are
less luminous in the central region than in the outskirts. In summary, disks in the 
core of clusters are less luminous, more compact and present a higher central surface 
brightness. The simulations performed by Moore et al. (1999) indicate that galaxy 
harassment is particularly strong for low surface brightness galaxies, which 
suggests that disks that can survive in the cluster core are 
the most compact ones, a scenario that is consistent with our results.

Finally, it is important to notice that for most of the analysis made in this work
the results clearly differ depending on whether redshift-confirmed members or  
galaxies in projection are considered. This indicates the importance of using 
redshift confirmed members.


\section{Acknowledgments}
This work was partially supported by the Consejo de Investigaciones 
Cient\'{\i}ficas y T\'ecnicas de la Rep\'ublica Argentina, CONICET; SeCyT, 
UNC, Agencia Nacional de Promoci\'on Cient\'{\i}fica and Agencia C\'ordoba 
Ciencia, Argentina. L. Infante anf H. Quintana acknowledge partial support 
from the Centro de Astrof\'{\i}sica FONDAP/CONICYT program.



\clearpage

\clearpage

\renewcommand{\arraystretch}{.6}

\begin{deluxetable}{cccccll}
\tablecaption{Observed clusters\label{table1}}
\tablewidth{0pc}
\tablehead{
\colhead{Abell Number} &
\colhead{$\alpha_{J2000}$} & 
\colhead{$\delta_{J2000}$} & 
\colhead{$\sigma$}&
\colhead{$v_r$}\\
&
\colhead{$^h\;\; ^m\;\;\; ^s\;\;\,$} & 
\colhead{$\degr\;\;\; \arcmin\;\;\;\arcsec\;$} & 
\colhead{[$kms^{-1}$]} &
\colhead{[$kms^{-1}$]}
}
\startdata
118  & 00 55 43.9 & $-$26 24 46 &  669 $\pm$ 127 & 34421 $\pm$ 159 \\ 
2734 & 00 11 20.1 & $-$28 52 52 &  784 $\pm$ 124 & 18502 $\pm$ 100 \\
2799 & 00 35 3.00 & $-$39 25 29 &  563 $\pm$ 62  & 19454 $\pm$ 127 \\ 
2800 & 00 37 58.7 & $-$25 05 30 &  335 $\pm$ 64  & 18943 $\pm$ 47 \\
2854 & 01 00 48.7 & $-$50 31 51 &  308 $\pm$ 44  & 18480 $\pm$ 51 \\
2923 & 01 32 18.0 & $-$31 05 36 &  670 $\pm$ 76  & 21420 $\pm$ 135 \\
2933 & 01 40 41.2 & $-$54 33 26 &  759 $\pm$ 72  & 27709 $\pm$ 105\\
3764 & 21 26 1.00 & $-$34 47 39 &  795 $\pm$ 123 & 22714 $\pm$ 110 \\
3809 & 21 49 51.7 & $-$43 52 55 &  560 $\pm$ 67  & 18785 $\pm$ 81 \\
3864 & 22 30 14.4 & $-$52 28 38 &  847 $\pm$ 188 & 30699 $\pm$ 161 \\
3915 & 22 47 37.0 & $-$52 03 09 &  815 $\pm$ 102 & 28925 $\pm$ 105 \\
3921 & 22 49 38.6 & $-$64 23 15 &  788 $\pm$ 111 & 27855 $\pm$ 105 \\
4010 & 23 31 10.3 & $-$36 30 26 &  743 $\pm$ 140 & 28766 $\pm$ 149 \\
4067 & 23 58 48.3 & $-$60 38 39 &  738 $\pm$ 442 & 29643 $\pm$ 181 \\
\enddata
\end{deluxetable}

\end{document}